\documentclass[apj]{emulateapj}
\usepackage[colorlinks,linkcolor={blue},citecolor={blue},urlcolor={red}]{hyperref}
\usepackage{epsfig}
\usepackage{graphicx}
\usepackage{natbib}
\usepackage{amsmath}
\usepackage{amsfonts}
\usepackage{amssymb}

\newcommand{\myemail}{\email{leech@shao.ac.cn}}

\slugcomment{Accepted for publication in ApJ} 
\shorttitle{Kinematics of groups of galaxies}
\shortauthors{Li et al.}

\begin{document}

\title{Internal kinematics of groups of galaxies in the Sloan Digital Sky Survey data release 7}

\author{Cheng  Li\altaffilmark{1}, Y. P.  Jing\altaffilmark{1}, Shude Mao\altaffilmark{2,3},
  Jiaxin Han\altaffilmark{1}, Qiuying Peng\altaffilmark{1}, Xiaohu Yang\altaffilmark{1}, \\
  H. J. Mo\altaffilmark{4}, Frank van den Bosch \altaffilmark{5}} 
\myemail

\altaffiltext{1}{Partner   Group   of  the   Max   Planck  Institute   for
  Astrophysics  at  the  Shanghai  Astronomical  Observatory  and  Key
  Laboratory for Research in Galaxies and Cosmology of Chinese Academy
  of   Sciences,    Nandan   Road   80,    Shanghai   200030,   China}

\altaffiltext{2}{National  Astronomical Observatories, Chinese  Academy of
  Sciences,   Beijing   100012,   China}

\altaffiltext{3}{Jodrell Bank Centre for Astrophysics, University of Manchester, 
Alan Turing Building, Manchester M13 9PL, UK}

\altaffiltext{4}{Department  of Astronomy,  University  of
  Massachusetts, Amherst MA 01003-9305, USA}

\altaffiltext{5}{Astronomy Department, Yale University, 
PO Box 208101, New Haven, CT 06520-8101, USA}

\begin{abstract}
We present  measurements of the velocity dispersion  profile (VDP) for
galaxy  groups in  the final  data release  of the  Sloan  Digital Sky
Survey   (SDSS).   For   groups  of   given  mass   we   estimate  the
redshift-space  cross-correlation  function (CCF)  with  respect to  a
reference  galaxy  sample,  $\xi^{(s)}(r_p,\pi)$, the  projected  CCF,
$w_p(r_p)$, and the real-space CCF, $\xi_{cg}(r)$. The VDP   is   then
extracted    from   the   redshift   distortion   in
$\xi^{(s)}(r_p,\pi)$,    by   comparing    $\xi^{(s)}(r_p,\pi)$   with
$\xi_{cg}(r)$.   We  find that  the  velocity  dispersion (VD)  within
virial radius ($R_{200}$) shows a  roughly flat profile, with a slight
increase  at radii below $\sim0.3 R_{200}$ for high  mass systems.  The
average VD within  the virial radius, $\sigma_v$,  is a strongly
increasing function of  central galaxy  mass.  We  apply the same
methodology   to  $N$-body  simulations   with  the  concordance
$\Lambda$  cold dark  matter  cosmology but  different  values of  the
density fluctuation  parameter $\sigma_8$, and we  compare the results
to the SDSS results.  We show that the $\sigma_v-M_\ast$ relation from
the  data  provides  stringent  constraints  on  both  $\sigma_8$  and
$\sigma_{ms}$, the dispersion in  $\log M_\ast$ of central galaxies at
fixed     halo    mass.      Our    best-fitting     model    suggests
$\sigma_8=0.86\pm0.03$  and  $\sigma_{ms}=0.16\pm0.03$.  The  slightly
higher value of  $\sigma_8$ compared to the WMAP7  result might be due
to a smaller matter density parameter assumed in our simulations.  Our
VD measurements also provide a  direct measure of the dark matter halo
mass  for central galaxies  of different  luminosities and  masses, in
good agreement with the results obtained by \citet{Mandelbaum-06} from
stacking the gravitational lensing signals of the SDSS galaxies.
\end{abstract}


\section{Introduction}
\label{sec:introduction}

Satellite  galaxies are  an  important tracer  of  the potential  well
within which they reside, as they can be observed to much larger radii
than other tracers. The potential  well is determined by both the dark
matter particles and baryons. In the past one and a half decade, $N$-body 
simulations provided excellent  understanding of the  mass profiles and  
shapes of dark matter  haloes. For example,  most haloes are well-fitted  by the
\citet[][hereafter NFW]{Navarro-Frenk-White-97} profile,  while their  shapes  are 
triaxial  ellipsoids \citep{Jing-Suto-02}. However, it is expected that
baryonic condensation at the centers  of   dark  matter  haloes  may  modify   
the  inner  profiles significantly,      particularly    on      the      
galaxy      scale \citep{White-Rees-78, Blumenthal-86}. Satellite  galaxies 
provide a valuable  probe of the
dark matter halo profiles and the baryonic condensation processes.

On  galaxy  scale, satellites  in  the  Milky  Way provides  the  best
dynamical  constraint  on  its  mass  profile out  to  about  200\,kpc
(\citealt{Kochanek-96, Wilkinson-Evans-99}).  
Stellar-dynamical analysis of massive early-type galaxies also 
provide the kinematics  of the  inner few  kpc, which  seems to  show  
the density profile  is roughly isothermal  (\citealt{Koopmans-06}).  
This  is much steeper  than the  inner slope  expected from  the NFW  
profile, which suggests that baryonic  processes  have  modified 
the  inner profiles, consistent with theoretical expectations
\citep[e.g.][]{Gnedin-04, Lin-06}.
For clusters  of galaxies,  using the CNOC1  survey \citet{Carlberg-97b}
showed that for their clusters,  the line of sight velocity rises from
0.1 virial  radius, reaches a peak  around 0.3 virial  radius and then
shows a  roughly flat  profile with a  very slight decline.  The line of  
sight velocity  dispersion profile  is consistent
with that derived by \citet{okas-Mamon-03}  from a detailed study of the nearby
Coma  cluster.  The  CNOC1 data  have  been used  by \citet{vanderMarel-00}  
to derive the mass distribution of clusters.

There have also been many studies that measure the {\em average} 
line-of-sight (los) velocity dispersion of the satellite galaxies within 
groups or clusters in order to infer the mass of their host dark matter 
halo. Early studies were mostly limited to rich galaxy clusters with a 
large number of satellite galaxies \citep[e.g.][]{Carlberg-96, Carlberg-Yee-Ellingson-97}.
For less massive systems, halo mass estimates are usually obtained for
central galaxies of similar luminosity or stellar mass by stacking the 
kinematics of their satellite galaxies \citep[e.g.][]{Erickson-Gottesman-Hunter-87,
Zaritsky-93, Zaritsky-White-94, Zaritsky-97}. Taking advantage of the
large redshift surveys, in particular the Two-degree Field
Galaxy Redshift Survey \citep[2dFGRS;][]{Colless-01} and the Sloan
Digital Sky Survey \citep[SDSS;][]{York-00}, recent studies have applied
this technique to large samples of satellite galaxies and studied the
dependence of the los velocity dispersion on galaxy properties such
as luminosity, stellar mass and optical color \citep[e.g.][]
{McKay-02, Brainerd-Specian-03, Prada-03, vandenBosch-04, Becker-07,
Conroy-07, Norberg-Frenk-Cole-08, More-09, More-11}.
These studies have well established that the los velocity dispersion 
increases with galaxy luminosity and mass, consistent with the theoretical
expectation that more massive galaxies are hosted by more massive halos.
In addition, \citet{More-09} found that the luminosity--halo mass relation
may differ with color, while the stellar mass--halo mass relation 
is less color-dependent.

In this work we measure the velocity dispersion profile of satellite 
galaxies in groups by modelling the redshift distortions in the two-point 
cross-correlation function between galaxies and groups of galaxies. 
It has been well established that, the two-point autocorrelation function 
(2PCF) of galaxies measured from redshift surveys is distorted along the 
line of sight due to the peculiar motions of galaxies. On small scales 
the 2PCF is stretched along the line of sight, called `Finger-of-God' 
(FoG) effect, and on large scales it is squashed due to the global 
infall of galaxies towards high-density regions \citep{Kaiser-87}. 
Thus the redshift distortion (RSD) in 2PCF contains useful information
about the relative motions of galaxies, and has been used to measure
the so-called pairwise velocity dispersion (PVD) of different classes
of galaxies by many authors \citep{Davis-Peebles-83, Mo-Jing-Borner-93, 
Fisher-94, Zurek-94, Marzke-95, Somerville-Davis-Primack-97, 
Jing-Mo-Boerner-98, Zehavi-02, Hawkins-03, Jing-Borner-04a, Li-06a}.
These measurements have been used to test/constrain both semi-analytic 
models of galaxy formation \citep[e.g.][]{Li-07} and halo occupation 
distribution models of galaxy distribution \citep[e.g.][]{Jing-Borner-04a,
Slosar-Seljak-Tasitsiomi-06, Tinker-07a, vandenBosch-07}, as well as
cosmological parameters \citep[e.g.][]{Jing-Mo-Boerner-98, Yang-04, 
Wang-08, Cabre-Gaztanaga-09}.

Here we use the same methodology as used in these previous studies, 
except that we use the galaxy-group  cross-correlation function to
probe the velocity distributions of galaxies with respect to dark 
matter haloes. When compared to the aforementioned `direct' measurements,
our method has the advantage that it is not influenced by the so-called
`interlopers', i.e. mis-identified group members \cite[see][ for detailed
discussion]{Yang-05a}. The cross-correlation functions between groups 
and galaxies have been determined by \citet{Yang-05c} based on an earlier
data release of the Sloan Digital Sky Survey \citep[SDSS;][]{York-00}.
In this study we make use of a sample of about 16,000 groups which 
is constructed from the final data release \citep[DR7;][]{Abazajian-09} 
of the SDSS using the halo-based group finding algorithm of \citet{Yang-07}. 
We estimate both the redshift-space cross-correlation function $\xi^{(s)}(r_p,\pi)$ 
and the real-space cross-correlation function $\xi_{cg}(r)$, between 
a given subsample of groups and a reference sample of galaxies selected 
from the SDSS. The velocity dispersion profile for each subsample is 
then derived by modelling the redshift distortion in $\xi^{(s)}(r_p,\pi)$.
As we will show, the velocity dispersion profiles within the virial
radius can be reliably determined using the current data over a 
wide range in central galaxy luminosity and mass. More interestingly,
comparisons with a set of high-resolution $N$-body simulations 
show that the average velocity dispersion within the virial radius 
as a function of central galaxy mass provides stringent constraints
on both the density fluctuation parameter $\sigma_8$, and the
correlation of the stellar mass of galaxies with the dark matter
mass of their host halos.

Throughout this paper we assume a cosmology model with the density
parameter $\Omega_m=0.27$ and the cosmological constant
$\Omega_\Lambda=0.73$, and a Hubble constant $H_0=100h
$kms$^{-1}$Mpc$^{-1}$ with $h=0.7$.

\begin{deluxetable*}{cccccc}
  \tablecaption{Subsamples of groups selected according to the stellar
    mass of their central galaxies, including the stellar mass range,
    the number of groups, the mean stellar mass and the mean
    luminosity of each subsample, as well as the one-dimensional
    velocity dispersion and halo mass determined for each subsample.}
  \tablewidth{0pt} 
  \tablehead{$\log(M_\ast/M_\odot)$ &
    $N_{group}$ & $\log(\langle M_\ast/M_\odot \rangle)$ &
    $\log(\langle L/L_\odot \rangle)$ & $\sigma_v/kms^{-1}$ &
    $\log(M_{h}/M_\odot)$ 
    \\ (1) & (2) & (3) & (4) & (5) & (6)}
  \startdata 
    $[10.3,10.6)$ & 913  & 10.49 & 10.33 & 133$\pm$11 & 12.17$\pm$0.13 
    \\ 
    $[10.4,10.7)$ & 1415 & 10.59 & 10.40 & 139$\pm$8  & 12.24$\pm$0.09
    \\
    $[10.5,10.8)$ & 2190 & 10.69 & 10.47 & 151$\pm$6  & 12.36$\pm$0.06 
    \\ 
    $[10.6,10.9)$ & 3145 & 10.78 & 10.53 & 166$\pm$5  & 12.50$\pm$0.05
    \\ 
    $[10.7,11.0)$ & 4245 & 10.88 & 10.61 & 188$\pm$6  & 12.69$\pm$0.05
    \\ 
    $[10.8,11.1)$ & 5257 & 10.97 & 10.69 & 216$\pm$7  & 12.88$\pm$0.04 
    \\ 
    $[10.9,11.2)$ & 6214 & 11.07 & 10.77 & 243$\pm$6  & 13.04$\pm$0.04
    \\ 
    $[11.0,11.3)$ & 6537 & 11.16 & 10.85 & 291$\pm$8  & 13.29$\pm$0.03
    \\ 
    $[11.1,11.4)$ & 6014 & 11.24 & 10.93 & 338$\pm$9  & 13.49$\pm$0.03 
    \\ 
    $[11.2,11.5)$ & 4685 & 11.33 & 11.01 & 385$\pm$10 & 13.65$\pm$0.03
    \\ 
    $[11.3,11.6)$ & 3115 & 11.42 & 11.09 & 439$\pm$15 & 13.82$\pm$0.04
    \\ 
    $[11.4,11.7)$ & 1806 & 11.51 & 11.18 & 522$\pm$22 & 14.04$\pm$0.05
    \\ 
    $[11.5,11.8)$ & 903  & 11.60 & 11.26 & 591$\pm$32 & 14.20$\pm$0.07
    \\ 
    $[11.6,11.9)$ & 374  & 11.69 & 11.35 & 654$\pm$38 & 14.32$\pm$0.07
    \enddata
    \label{tbl:samples}
\end{deluxetable*}

\section{Data}
\label{sec:data}

\subsection{Group catalog}
\label{sec:group_catalog}

The galaxy group catalog used in this paper is constructed by
\citet{Yang-07} from {\tt sample dr72} of the New York University
Value-Added Galaxy Catalogue (NYU-VAGC).  The NYU-VAGC is a catalogue
of local galaxies (mostly below $z\approx0.3$) selected from the SDSS
data release 7 \citep[DR7;][]{Abazajian-09}, publicly available at
http://sdss.physics.nyu.edu/vagc/, and is described in detail in
\citet{Blanton-05a}.  The main virtue of the NYU-VAGC is that it
provides a detailed account of the selection effects in the survey,
thus suitable for statistical studies of the galaxy distribution and
large-scale structure in the local Universe.

To select the groups of galaxies, a modified version of the halo-based
group-finding algorithm developed in \citet{Yang-05a} is applied to a
sample of $\sim6.4\times10^5$ galaxies selected from the NYU-VAGC {\tt
  sample dr72} with redshifts in the range $0.01\le z\le 0.20$ and
with a redshift completeness above 70\%. The reader is referred to
\citet{Yang-05a} and \citet{Yang-07} for detailed description of the
group finder.  The group catalogue contains about half a million
systems, of which the majority have only a single member. In this work
we use a subset of $\sim$16,000 groups that have at least three member
galaxies.

We use the most massive galaxy member of each group as the group
center which is called the ``central'' galaxy in what follows.  The
stellar mass of each central galaxy accompanies the NYU-VAGC release,
which is estimated based on its redshift and the five-band magnitudes 
from SDSS photometric data, as described in detail in 
\citet{Blanton-Roweis-07}.
This estimate corrects implicitly for dust and assumes a universal
stellar initial mass function (IMF) of \citet{Chabrier-03} form.  As
demonstrated in Appendix A of \citet{Li-White-09}, once all estimates
are adapted to assume the same IMF, the Blanton \& Roweis masses agree
quite well with those obtained from the simple, single-color
estimator of \citet{Bell-03} and also with those derived by
\citet{Kauffmann-03} from a combination of SDSS photometry and
spectroscopy. In this work we use the ``total masses'' instead of 
the ``Petrosian masses'' used by \citet{Li-White-09}, obtained by
correcting the latter using SDSS ``model magnitudes'' 
\citep[see Appendix A of][for details]{Guo-10}.

From the group catalogue, we select 18 subsamples according to the
stellar mass of the central galaxies, ranging from
$\log (M_\ast/M_\odot)=$10.0 to 12.0.  Each subsample includes
central galaxies in a stellar mass interval of 0.3 dex, with
successive subsamples overlapping by 0.2 dex. As we will show below,
the velocity dispersion profile cannot be reliably determined for 
subsamples below $\log (M_\ast/M_\odot)=10.3$. We will also ignore
the most massive subsample with $11.9\le\log (M_\ast/M_\odot)<12.0$
when measuring the velocity dispersion profile, because its redshift-space
cross-correlation function is noisy due to the small sample size.
The stellar mass range, the number of groups, and the mean stellar mass 
and luminosity of the rest 14 subsamples are listed in the first four 
columns in Table 1. The luminosity of each galaxy is computed from its 
$r-$band absolute magnitude and the absolute magnitude of the Sun is 
assumed to be 4.76 mag following \citet{Blanton-03c}.

\subsection{Reference galaxy sample and random sample}
\label{sec:reference_sample}

We have  constructed a magnitude-limited  galaxy sample from  the 
NYU-VAGC {\tt sample dr72},
which will serve as our {\em reference} sample when measuring the
group-galaxy cross-correlation functions.  This consists of about half
a million galaxies  with $r<17.6$, $-24<M_{^{0.1}r}<-16$ and redshifts
in  the  range $0.01<z<0.2$.   Here,  $r$  is  the $r$-band  Petrosian
apparent   magnitude,   corrected   for   Galactic   extinction,   and
$M_{^{0.1}r}$ is the  $r$-band Petrosian absolute magnitude, corrected
for  evolution  and $K$-corrected  to  its  value  at $z=0.1$.   These
selection criteria, with the exception  of the redshift range, are the
same  as in our  previous papers  where we  studied the  clustering of
galaxy luminosity and stellar mass \citep[e.g.][]{Li-12a}.

To obtain reliable estimates of the cross-correlation functions, which 
will form the basis of our study, the cross pair counts between the
groups and the reference galaxies must be compared with the cross pair 
counts between the same set of groups and a `random sample'. The random
sample is unclustered but fills the same region of the sky and has the same
position- and redshift-dependent selection effects as the reference
sample. We have constructed our random sample from the observed reference
sample itself, as described in detail in \citet{Li-06c}. For each
real galaxy we generate 10 sky positions at random within the mask of 
{\tt sample dr72}, and we assign to each of them the redshift of the
real galaxy. The resulting random sample is valid for clustering
analyses, provided that the survey area is large enough so that structures 
in the real sample are wiped out by randomizing in angle, and that 
the effective depth of the survey does not vary from region to region.
Both are true to good accuracy for our sample, which covers $\ga 6000$ deg$^2$,
is complete down to $r=17.6$ and is little affected by foreground dust
over the entire survey region. Extensive tests show that random samples
constructed in this way produce indistinguishable results from those
using the traditional method \citep{Li-06c}. 

\subsection{Simulations and dark matter subhalo catalogs}
\label{sec:simulations}

In this paper, we use three $N$-body simulations and dark matter
halo catalogs constructed from them to study the connections of galaxy
velocity dispersions with halo masses and cosmological parameters. 
The simulations use $1024^3$ particles to follow the dark-matter distribution 
in a cubic region with $300 h^{-1}$Mpc on a side, corresponding to
a particle mass of $1.87\times10^9h^{-1}$M$_\odot$. These are obtained 
using an upgraded version of the particle-particle-particle-mesh (P$^3$M) 
code of \citet{Jing-Suto-98, Jing-Suto-02}, which has incorporated the 
multiple-level P$^3$M gravity solver for high-density regions 
\citep{Jing-Suto-00, Jing-Suto-Mo-07}. The cosmology model assumed is a
spatially flat $\Lambda$CDM model with the same parameters $\Omega_m=0.268$, 
$\Omega_\Lambda=0.732$, $\Omega_b=0.045$, $H_0=71$kms$^{-1}$Mpc$^{-1}$ 
and $n=1$, except that a different density fluctuation parameter is adopted, 
which is $\sigma_8=0.75$, 0.85 and $0.95$, for the three simulations 
respectively. The simulations are described in detail in \citet{Jing-Suto-Mo-07}.
The cosmology for the simulations is nearly identical to the cosmology 
we assume in this study (mentioned in the introduction).

Dark matter halos are identified for each of the simulation outputs
using the standard Friends-of-Friends (FoF) algorithm with the linking
length set to be 0.2 times the mean particle separation \citep{Davis-85}. 
The Hierarchical Bound-Tracing (HBT) algorithm recently developed by 
\citet{Han-11} is then applied to split the FoF halos into disjoint, 
self-bound subhalos. The HBT algorithm finds and traces dark matter subhalos 
in simulations based on their merger hierarchy. Thus, when compared to 
previously existing subhalo finders, this new algorithm has the unique 
advantage that it is able to well resolve the subhalos in high density 
environment, while keeping strict physical track of their merger history. 
Inside each FoF halo, the subhalos are divided into two classes which
are treated differently during tracing: `central' subhalo and `satellite'
subhalos. The central subhalo is defined as the most massive subhalo
of an FoF halo, is able to grow through accreting mass within the host
halo and normally contains most of its mass. All the subhalos in an
FoF halo except the central one are called satellite subhalos.

We determine the position and velocity of each subhalo to be the center 
of mass and bulk velocity of its core, consisting of 25\% of its bound 
particles with the lowest potential. For each central subhalo, we define 
the virial radius, $R_{200}$, as the radius within which the mean density 
of dark matter particles is equal to 200 times the critical value. The 
virial mass of the central subhalo is then defined as the mass within 
$R_{200}$:
\begin{equation}\label{eqn:m200}
M_{200} = 200\rho_{crit}\frac{4\pi}{3}R_{200}^3=100G^{-1}H^2(z)R_{200}^3.
\end{equation}
For a satellite subhalo, we use the $M_{200}$ at the epoch when it  
was last the central subhalo in its host halo. Hereafter, we refer
to both central and satellite subhalos as `halos', and we refer to 
$M_{200}$ defined in this way as the `halo mass', if not specifically
pointed out.

\section{Results}

\subsection{Group-galaxy cross-correlation functions}
\label{sec:correlation_functions}

In this work we estimate the velocity dispersion of galaxies by modelling 
the redshift distortions in group-galaxy cross-correlation functions. Thus, 
for each subsample of groups listed in Table 1, we begin by estimating
the redshift-space cross-correlation function, $\xi^{(s)}(r_p,\pi)$, 
between the group central galaxies and the galaxies in the reference
sample, using the commonly-used estimator
\begin{equation}
  \xi^{(s)}(r_p,\pi) = \frac{N_R}{N_G}\frac{CG(r_p,\pi)}{CR(r_p,\pi)}-1.
\end{equation}
Here $r_p$ and $\pi$ are the separations perpendicular and parallel to 
the line of sight; $N_G$ and $N_R$ are the number of galaxies in the 
reference sample and in the random sample; $CG(r_p,\pi)$ and $CR(r_p,\pi)$ 
are the cross pair counts between the group subsample and the reference 
sample, and between the group subsample and the random sample respectively,
both with perpendicular separations in the bins 
$\log r_p\pm\frac{1}{2}\Delta\log r_p$ and with radial separations in the 
bins $\pi\pm\frac{1}{2}\Delta\pi$. To reduce sampling noise, the random sample is
constructed with a number of particles that is 10 times the number of 
galaxies in the reference sample, i.e., $N_R=10\times N_G$.

An estimate of the projected cross-correlation function, $w_p(r_p)$, is then 
obtained by integrating $\xi^{(s)}(r_p,\pi)$ over the line-of-sight 
separations $\pi$. We take
\begin{equation}
  w_p(r_p) = \int_{-\pi_{\rm max}}^{\pi_{\rm max}} \xi^{(s)}(r_p,\pi)d\pi 
  = \sum_{i}\xi^{(s)}(r_p,\pi_i)\Delta\pi_i,
\end{equation}
where we choose $\pi_{\rm max}=40 h^{-1}$Mpc as the outer limit for the 
integration depth (in order to limit noise from distant uncorrelated regions) 
so that the summation for computing $w_p(r_p)$ runs from 
$\pi_1=-39.5 h^{-1}$Mpc to $\pi_{80}=39.5 h^{-1}$Mpc, given that we use 
bins of width $\Delta\pi_i=1 h^{-1}$Mpc. We note that we have tried with
different outer limits for the integration depth and the resulting $w_p(r_p)$
changes little. We have corrected the effect of fiber collisions in the
SDSS data using the method described in \citet{Li-06b}.

\begin{figure*}
  \begin{center}
    \epsfig{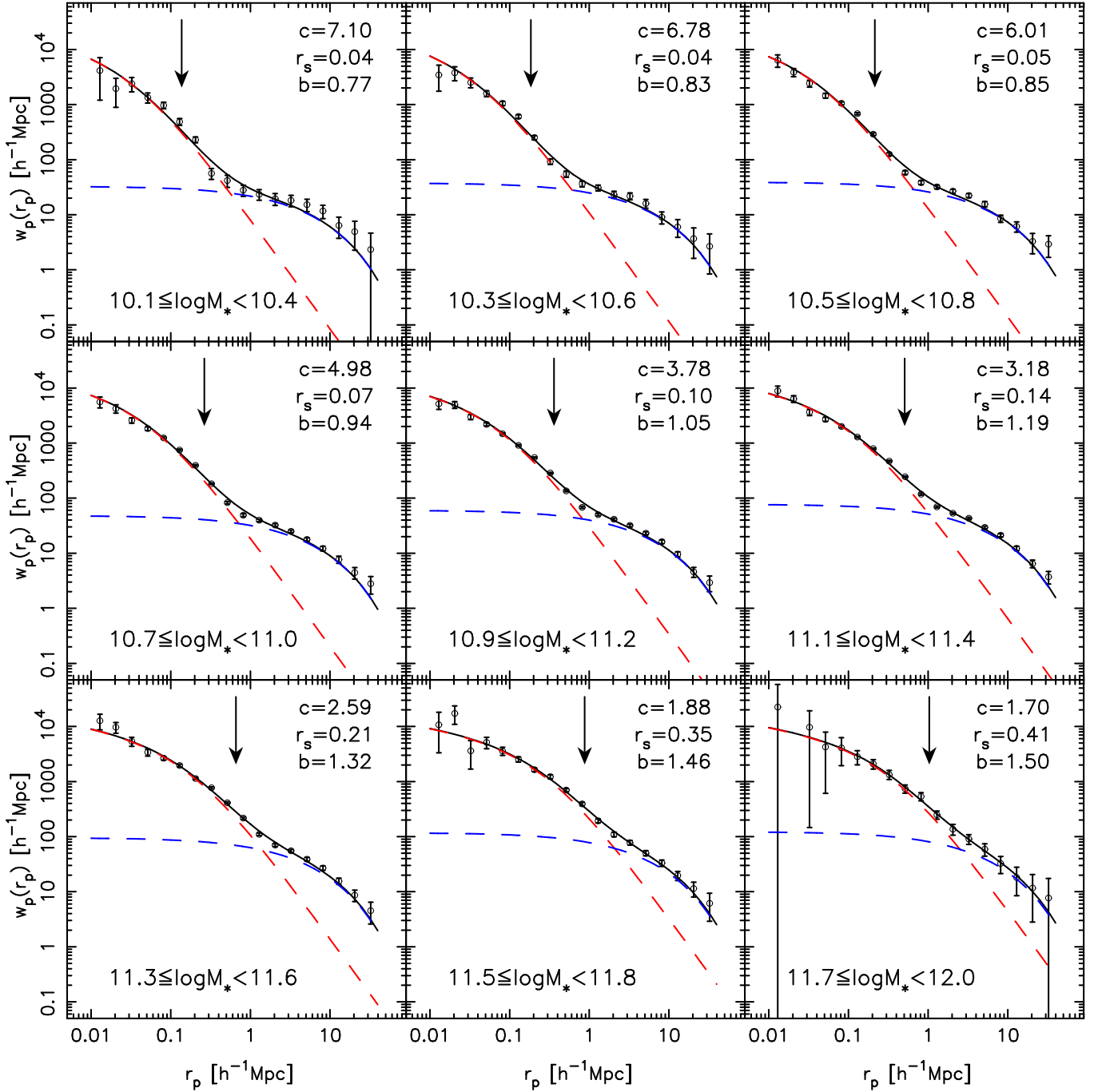}
  \end{center}
    \caption{The projected cross-correlation function between the
      central galaxies of groups and the reference galaxies, 
      estimated for some of the group subsamples listed in
      Table 1. 
      The symbols with error bars are measured using the SDSS
      DR7 group catalogue. The solid lines are 
      the best-fits of Eq.(\ref{eqn:wrpmodel}) to the data. The dashed and dotted 
      lines are the small-scale component modelled by an 
      NFW profile and the large-scale component that is a biased version of 
      the projected linear two-point correlation function of dark matter 
      in a WMAP7 cosmology. The stellar mass range of the group central
      galaxies and the best-fitting values of the model
      parameters are indicated in each panel (note that the parameter
      $r_s$ is in units of $h^{-1}$Mpc). 
      The virial radius $R_{200}$ given by the estimated halo masses
      listed in Table 1 is indicated by an vertical arrow in every panel.
      See the text for details.}
  \label{fig:wrp}
\end{figure*}

In Fig.~\ref{fig:wrp}, we show the projected cross-correlation functions
determined in this way for some of our group subsamples. 
The errors on the cross-correlation function measurements are estimated 
using the bootstrap resampling technique \citep{Barrow-Bhavsar-Sonoda-84}. 
We generate 100 bootstrap samples from the group subsample and compute the 
cross-correlation functions for each sample using the weighting scheme 
(but not the approximate formula) given by \citet{Mo-Jing-Boerner-92}. 
The errors are then given by the $1\sigma$ scatter of the measurements 
among these bootstrap samples. 

In all the cases the projected cross-correlation function
can well be separated into two parts: a steeper inner part at separations
below around 1 $h^{-1}$Mpc and a flat outer part at larger separations.
In the language of the halo model \citep[e.g.][]{Cooray-Sheth-02}, the
inner part is dominated by the `one-halo' term where the pair counts are
mostly galaxy pairs in the same halo (central-satellite pairs in the case
of cross-correlation functions as in this study), and the outer part is 
dominated by the `two-halo' term where galaxy pairs are mostly in separate 
halos (central-central plus central-satellite pairs between different 
groups). The separation where the transition between the one-halo and 
two-halo terms occurs increases monotonically from $\sim$200 $h^{-1}$kpc 
for the lowest-mass systems in our sample with $M_\ast\sim10^{10}M_\odot$, 
up to $\sim$1 $h^{-1}$Mpc for the most massive systems with 
$M_\ast\sim10^{12}M_\odot$. The one-halo term shows 
a systematic change in shape with the mass of central galaxies, with 
steeper slopes at lower masses and flatter slopes at higher masses.
This implies that the distribution of satellite galaxies in low-mass
halos is more concentrated than that in high-mass halos. Moreover, the
amplitude of the two-halo term increases with increasing mass, reflecting
the tight correlation 
between the stellar mass of the central galaxies and the mass of their 
host dark matter halos. In contrast, the slope of the two-halo term seems 
to vary little, reflecting the known fact that 
the bias in the galaxy distribution is related in a very simple way to 
the bias in the distribution of dark haloes. The features seen in the
group-galaxy cross-correlation functions are all well consistent with
what we found in \citet{Yang-05c}, but with smaller error bars thanks
to the improvement of the data.

In the plane-parallel approximation, the projected cross-correlation 
function $w_p(r_p)$ is expected to be directly related to the real-space 
cross-correlation function $\xi_{cg}(r)$ by an Abel transform,
\begin{equation}\label{eqn:wrpmodel}
w_p(r_p) = \int_{-\infty}^\infty \xi_{cg}\left(\sqrt{r_p^2+y^2}\right)dy
       = 2\int_{r_p}^\infty \frac{r\xi_{cg}(r)}{\sqrt{r^2-r_p^2}} dr.
\end{equation}
Considering the one- and two-halo term features seen in the $w_p(r_p)$
measurements as discussed above, as well as previous theoretical studies 
of the cross-correlation function between halo centers and dark matter
particles \citep[e.g.][]{Hayashi-White-08}, we model the real-space 
cross-correlation function $\xi_{cg}(r)$ as a linear combination of two 
components as follows, 
\begin{equation}\label{eqn:xirmodel}
  \xi_{cg}(r) = \xi_{\rm NFW}(r) + b\xi_{dm}(r).
\end{equation}
The first component, $\xi_{\rm NFW}(r)$, is the NFW profile,
\begin{equation}\label{eqn:nfw}
\xi_{\rm NFW}(r) = \frac{\delta_c}{(r/r_s)(1+r/r_s)^2},
\end{equation}
where
\begin{equation}
  \delta_c = \frac{200}{3\Omega_m}\frac{c^3}{\ln(1+c)-c/(1+c)}.
\end{equation}
The second component, $\xi_{dm}(r)$, is the linear two-point correlation
function of dark matter at $z=0$ for the WMAP7 cosmology, obtained by
Fourier transforming $P_l(k)$, the linear power spectrum calculated using
the {\tt CAMB} code of \citet{Lewis-Challinor-Lasenby-00}. 
Thus, the model has three free parameters: $c$, $r_s$ and $b$, and we
will call them concentration parameter, characteristic radius and bias 
factor in what follows, though they may have different meanings from their 
original definitions in the literature. 

\begin{figure*}
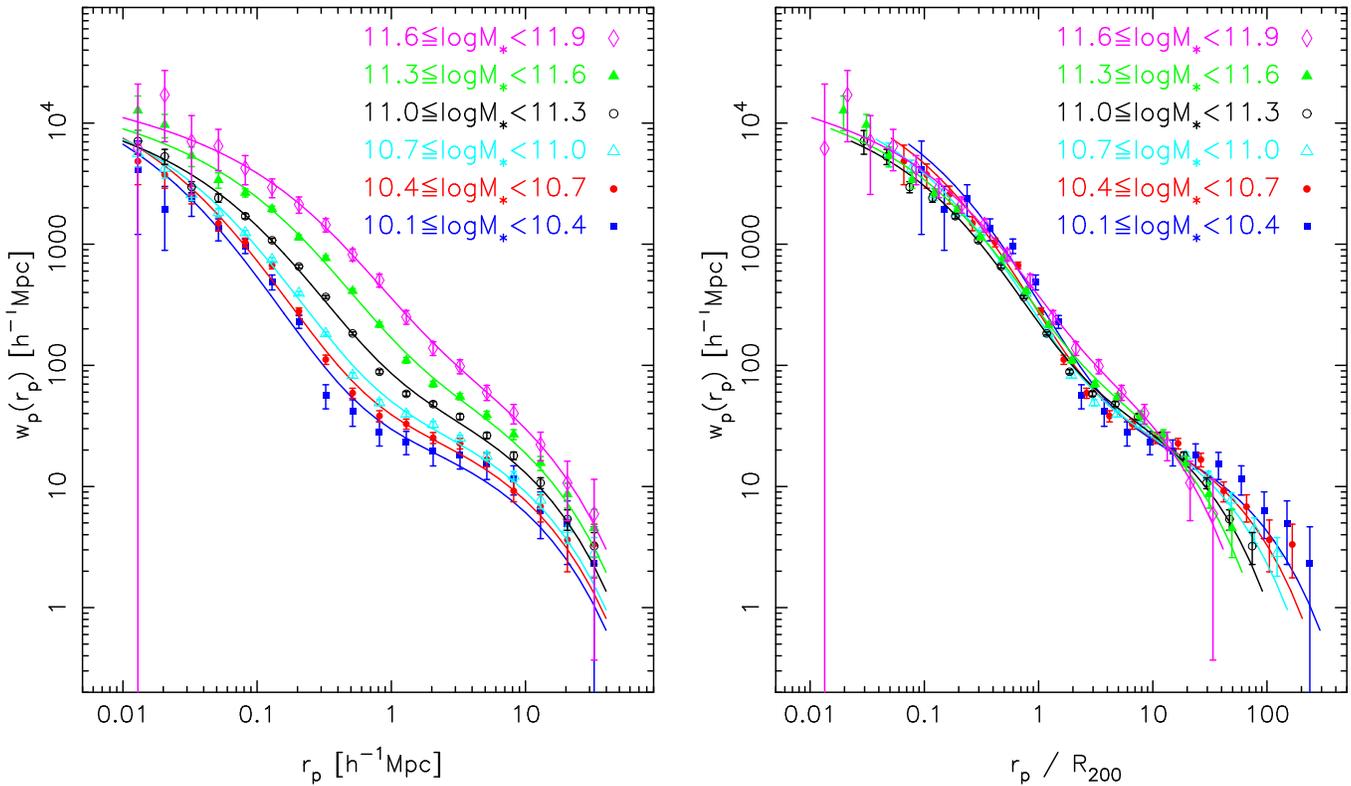

  \begin{center}
    \epsfig{figure=f2a.ps,width=0.48\hsize}
    \hspace{0.02\hsize}
    \epsfig{figure=f2b.ps,width=0.48\hsize}
  \end{center}
  \caption{{\it Left}: the $w_p(r_p)$ measurements from the
SDSS/DR7 group catalogue are plotted in symbols with error bars
for some of our group subsamples, and are compared to the
best-fits of Eqn.(\ref{eqn:wrpmodel}) to the data which are
plotted in solid lines. {\it Right}: same as in the left panel
except that the projected separation $r_p$ is scaled by the
virial radius $R_{200}$ on the x-axis. The stellar mass ranges
of the central galaxies are indicated.}
  \label{fig:wrp_1panel}
\end{figure*}

For a given set of parameters, the model predicts both the real-space 
correlation function $\xi_{cg}(r)$ through Eqn.~(\ref{eqn:xirmodel}) and 
the projected function $w_p(r_p)$ through Eqn.~(\ref{eqn:wrpmodel}).
We determine a best-fitting model for each of our group subsamples by 
comparing the predicted $w_p(r_p)$ with the observed one, and we define the
best-fitting model to be the one giving a minimum $\chi^2$.
Our best models are plotted in Fig.~\ref{fig:wrp} 
as solid black lines, and the two components of each model are plotted 
in red and blue dashed lines separately. The model parameters are also 
indicated. As can be seen, the two-component model provides a compact
and accurate description of the observed $w_p(r_p)$ for all our group 
subsamples. It is interesting that all the model parameters show systematic 
trends when one goes from the lowest-mass groups to the highest-mass groups,
in the sense that both $r_s$ and $b$ increase with mass, while 
the concentration parameter, $c$, decreases with mass. This is again
consistent with the picture that more massive galaxies are 
hosted by more massive dark halos within which the galaxy distribution 
is less concentrated and extends to a larger radius. 

Moreover, the concentration parameter of satellite distribution as inferred 
from our best-fitting models is smaller than the concentration of dark matter 
halos given by high-resolution simulations \citep[e.g.][]{Zhao-09}, 
implying that the distribution of galaxies in dark halos is more extended than 
dark matter particles. This is consistent with some previous studies
\citep[e.g.][]{Yang-05c, Chen-06, More-vandenBosch-Cacciato-09}, though
other studies produced opposite results, especially for luminous
red galaxies \citep[e.g.][]{Masjedi-06, Watson-10, Watson-12, 
Tal-Wake-vanDokkum-12}. Nevertheless, we would like to emphasize that 
our model parameters should not be overinterpreted, although they exhibit 
reasonable values and interesting trends with galaxy mass that seem to be 
easily understood from our knowledge of galaxy-halo connections.

In Fig.~\ref{fig:wrp_1panel} we plot again the measurements 
and the best-fits of the projected cross-correlation function, 
as functions of both $r_p$ (left panel) and $r_p/R_{200}$ (right panel). 
The systematic trends of $w_p(r_p)$ with central galaxy mass as 
seen from Fig.~\ref{fig:wrp} are more clearly seen here. These include 
the increase in the amplitude of the two-halo term on large scales, 
the change in the shape of the one-halo term on small scales, 
and the shift in the transition scale between the two terms on intermediate scales. 

\subsection{Velocity dispersion profiles}
\label{sec:velocity_dispersions}

\begin{figure*}
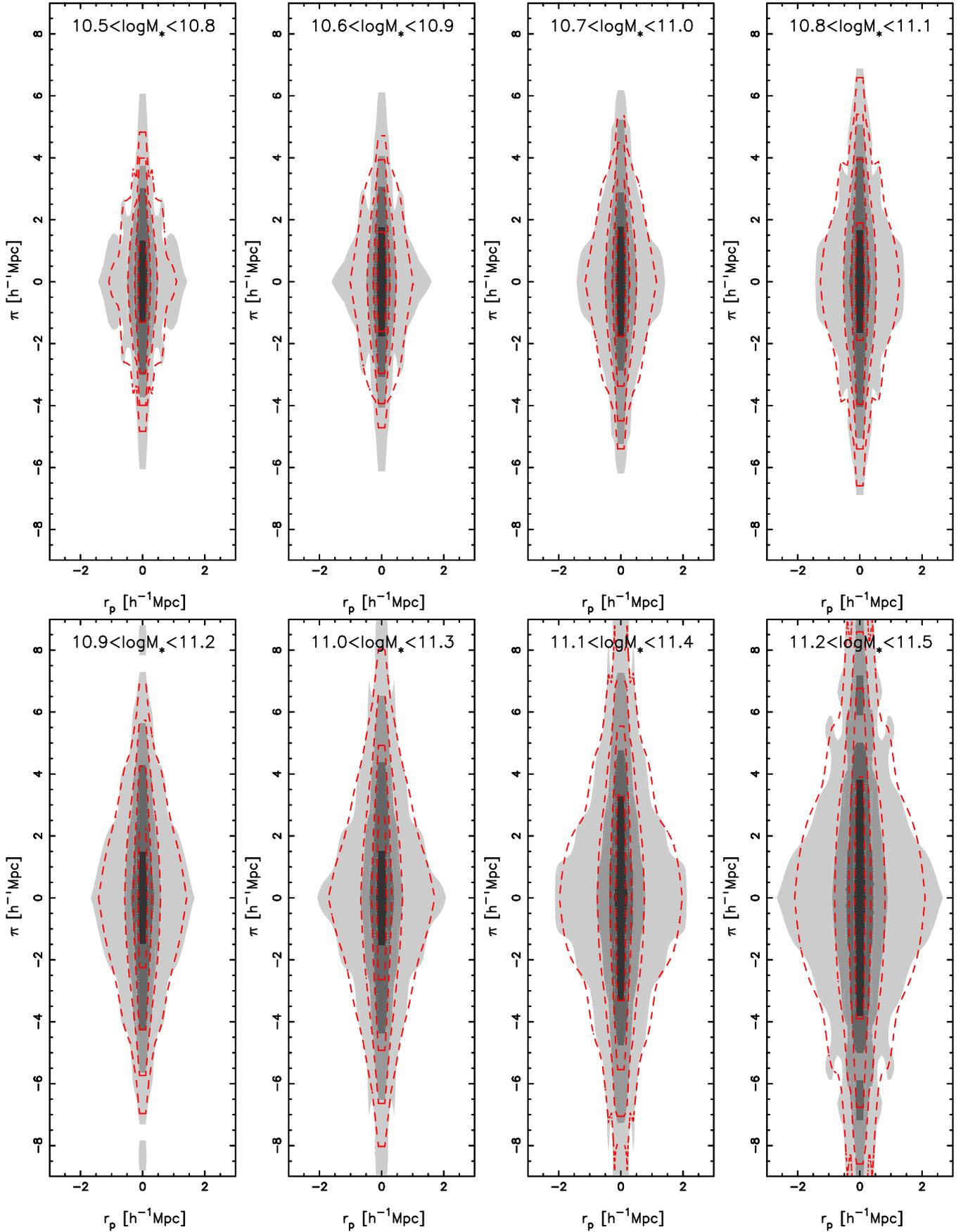

  \begin{center}
    \epsfig{figure=f3a.ps,width=0.245\hsize}
    \epsfig{figure=f3b.ps,width=0.245\hsize}
    \epsfig{figure=f3c.ps,width=0.245\hsize}
    \epsfig{figure=f3d.ps,width=0.245\hsize}
    \epsfig{figure=f3e.ps,width=0.245\hsize}
    \epsfig{figure=f3f.ps,width=0.245\hsize}
    \epsfig{figure=f3g.ps,width=0.245\hsize}
    \epsfig{figure=f3h.ps,width=0.245\hsize}
  \end{center}
  \caption{The redshift-space group-galaxy cross-correlation function
$\xi^{(s)}(r_p,\pi)$ for groups with central galaxies in different
stellar mass intervals, as indicated in each panel. The gray
shaded regions show the measurements from the SDSS/DR7 data and
the dashed red lines are the best-fits of Eqn.(\ref{eqn:xipvmodel})
to the data.}
  \label{fig:xipv}
\end{figure*}

In the previous subsection we have obtained the group-galaxy cross-correlation 
functions, both in redshift space, $\xi^{(s)}(r_p,\pi)$ and 
in real space, $\xi_{cg}(r)$. In Fig.~\ref{fig:xipv} we show the contours
of $\xi^{(s)}(r_p,\pi)$ for some of our group subsamples. Both the 
`Finger-of-God' 
(FoG) effect on small scales and the infall effect \citep{Kaiser-87} on 
large scales are clearly visible: on small scales $\xi^{(s)}(r_p,\pi)$ is 
stretched in the $\pi$-direction and on large scales the contours are squashed. 
The FoG effect is caused by the peculiar, virialized motions of galaxies 
within dark matter haloes, and is much more pronounced in the more massive 
haloes, reflecting their larger velocity dispersions. Here, we model these 
redshift-space distortions in detail, in order to infer the velocity field 
in and around dark haloes.

Our method relies on the fact that the peculiar motions of galaxies affect 
only their radial distances in redshift space. Thus the information for 
peculiar velocities along the line of sight can be recovered by modelling 
the redshift-space cross-correlation function $\xi^{(s)}(r_p,\pi)$ as a 
convolution of the real-space function $\xi_{cg}(r)$ with the distribution function
of the peculiar velocity $f(v_{cs})$:
\begin{equation}\label{eqn:xipvmodel}
  \xi^{(s)}(r_p,\pi) = \int f(v_{cs})\xi_{cg}\left(\sqrt{r_p^2+(\pi-v_{cs})^2}\right)dv_{cs},
\end{equation}
where $v_{cs}$ is the peculiar velocity of satellite galaxies relative
to the central galaxy. We adopt a Gaussian form for $f(v_{cs})$:
\begin{equation}
  f(v_{cs}) = \frac{1}{\sqrt{2\pi}\sigma_{v}}
  \exp\left[\frac{-(v_{cs}-\overline{v_{cs}})^2}{2\sigma_v^2}\right],
\end{equation}
where $\overline{v_{cs}}$ is the mean and $\sigma_v$ is the dispersion of the 
one dimensional peculiar velocities. Note that we have assumed that the 
distributions of the peculiar velocity and the velocity dispersion are isotropic.
Assuming an infall model for $\overline{v_{cs}}(r)$, the velocity dispersion $\sigma_v$ 
can then be estimated as a function of the projected separation $r_p$ by comparing 
the observed $\xi^{(s)}(r_p,\pi)$ with the modelled one.

Following \citet{Croft-Dalton-Efstathiou-99}, we use the non-linear spherical 
collapse model as our infall model,
\begin{equation}\label{eqn:infallmodel}
  \overline{v_{cs}}(r) = v_{l}(r) 
           \left[1+\delta(r)\right]^{-0.25}
           \exp\left[-\frac{\delta(r)}{\delta_{cut}}\right].
\end{equation}
The first factor at the right-hand side, $v_{l}$, is the infall velocity in the 
linear perturbation case, which depends on the distance to the local density 
maximum and is directly related to the density contrast \citep{Peebles-80}:
\begin{equation}\label{eqn:vlin}
  v_{l}(r) = -\frac{1}{3}H_0\Omega_m^{0.6}r\delta(r),
\end{equation}
where $\delta(r)$ is the overdensity inside radius $r$,
\begin{equation}
  \delta(r) = \frac{3J_3^{c\rho}(r)}{r^3}, \hspace{0.4cm}
  J_3^{c\rho}(r) = \int_0^{r}\xi_{c\rho}(x)x^2dx.
\end{equation}
Here $\xi_{c\rho}$ is the linear cross-correlation between 
galaxy clusters (or groups in our case) and matter. We use the simple linear 
biasing picture, so that 
\begin{equation}
  J_3^{c\rho}(r) = \frac{J_3^{cg}(r)}{b} = \frac{1}{b}
  \int_{0}^{r}\xi_{cg}(x)x^2dx,
\end{equation}
where $\xi_{cg}$ is the group-galaxy cross-correlation function
in real-space and $b$ is the linear bias factor in Eqn.(\ref{eqn:xirmodel}), 
both of which have been obtained in the previous subsection for each of our 
group subsamples. 
The power-law part at the right-hand side in Eqn.(\ref{eqn:infallmodel}) 
is introduced by \citet{Yahil-85} (also used by \citealt{Lilje-Efstathiou-89}) 
as a good approximation to the exact solution for the non-linear
collapse of spherically symmetric clusters. The exponential 
truncation in Eqn.(\ref{eqn:infallmodel}) is suggested by 
\citet{Croft-Dalton-Efstathiou-99}
in order to take into account the highly non-linear behavior of the infall 
velocity at high overdensities. Tests against N-body simulations 
showed that Eqn.(\ref{eqn:infallmodel}) with $\delta_{cut}=50$ 
provides an accurate description of the average infall velocity of 
matter into galaxy clusters \citep{Croft-Dalton-Efstathiou-99}.

\begin{figure*}
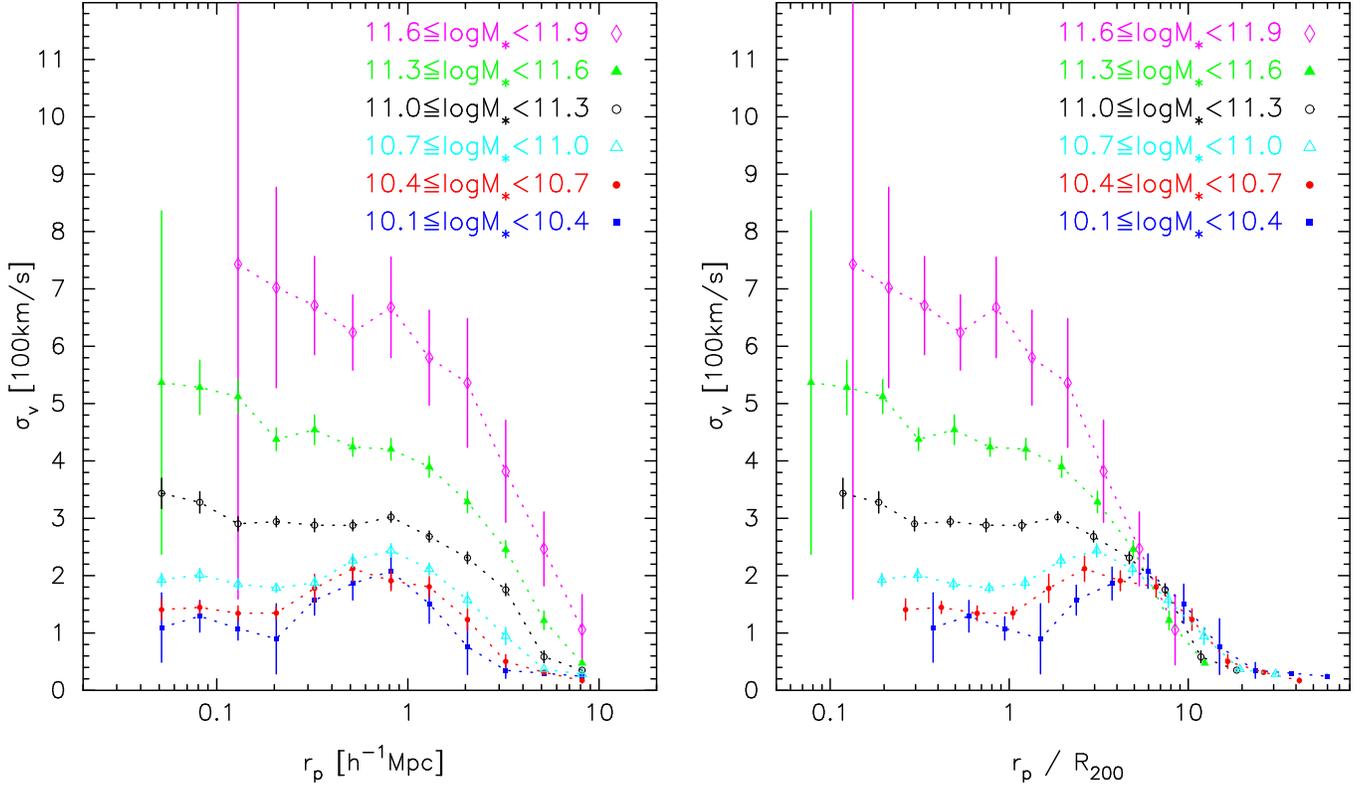

  \begin{center}
    \epsfig{figure=f4a.ps,width=0.48\hsize}
    \hspace{0.02\hsize}
    \epsfig{figure=f4b.ps,width=0.48\hsize}
  \end{center}
  \caption{{\it Left}: the velocity dispersion profile for some 
of our group subsamples, obtained by applying Eqn.(\ref{eqn:xipvmodel}) 
to the data. {\it Right}: same as the left panel except that 
the projected separation $r_p$ is scaled by the virial radius
$R_{200}$ on the x-axis. The stellar mass ranges of the group 
subsamples are indicated in both panels.}
  \label{fig:profiles}
\end{figure*}

\begin{figure*}
  \begin{center}
    \epsfig{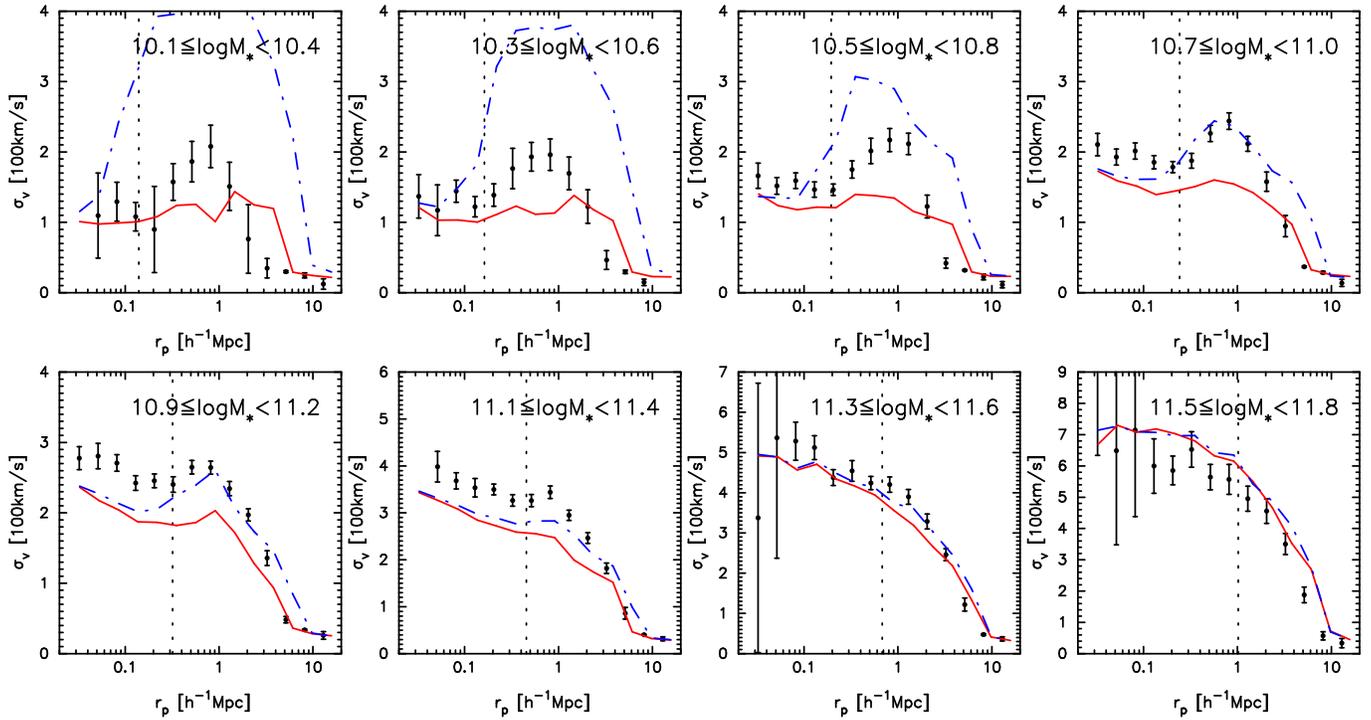}
  \end{center}
  \caption{The velocity dispersion profile measured from the
SDSS data are plotted in black symbols with error bars for some
of our group subsamples, and are compared with the measurements
from dark matter subhalo catalogs before (blue dash-dotted lines) and
after (red solid lines) `blending' close halos
(see the text for details). The vertical lines indicate the
virial radii corresponding to the halo masses listed in Table 1.}
  \label{fig:sigma_6127}
\end{figure*}

In Fig.~\ref{fig:xipv} we plot the contours of our best-fitting models
of $\xi^{(s)}(r_p,\pi)$ obtained by applying Eqn.(\ref{eqn:xipvmodel})
to the data. In Fig.~\ref{fig:profiles} we plot the velocity dispersion,
$\sigma_v$, obtained by applying Eqn.(\ref{eqn:xipvmodel}) to 
the data, as functions of both the projected separation ($r_p$, left 
panel) and the separation scaled by the virial radius ($r_p/R_{200}$,
right panel). There are several interesting trends/features
that can be read from this figure. First, at fixed radii the velocity 
dispersion is a strongly increasing function of central galaxy mass
(thus halo mass), and the mass dependence becomes stronger for
more massive systems. The velocity dispersion within the scales of
the one-halo term ranges from $\sim100$ kms$^{-1}$ for the lowest-mass 
systems in our sample ($M_\ast\sim 2\times 10^{10}M_\odot$), up to 
$\sim 700$ kms$^{-1}$ for the most massive systems with $M_\ast\sim5\times10^{11}M_\odot$.
Second, the velocity dispersion shows a roughly flat profile at radii 
smaller than a few 100 $h^{-1}$kpc, with a slight increase at the 
smallest radii for the massive systems with $M_\ast\ga 3\times10^{11}M_\odot$. 
This echoes the early results revealed from CNOC1 survey by 
\citet{Carlberg-97b} who found that for their clusters the line-of-sight
velocity showed a flat profile at radii larger than 0.3 virial radius
with a very slight decline \citep[see also][]{Carlberg-96, Biviano-Katgert-04},
as well as later studies on galaxy groups by \citet{Carlberg-01} who
found flat or slowly rising profiles for small groups. Finally, the velocity 
dispersion profile exhibits a broad bump at around 1 Mpc, which is more 
pronounced for less massive systems. As we will show below, this is 
likely caused by contamination from neighboring larger halos where the 
high-speed galaxies contribute significantly to the velocity dispersion 
measurement of the smaller system.

\subsection{Understanding the origin of the bump at around 1 Mpc}
\label{sec:the_bump}

In order to understand the origin of the bump at around 1 Mpc
in the velocity dispersion profile of the low-mass systems,
we have performed the same analyses using our simulations and 
compared the results with the observation obtained above.

We take all the halos from the $z=0$ outputs of our simulations, and 
we assign a stellar mass to each halo by applying the commonly-applied 
subhalo abundance matching model \citep[SHAM;][]{Vale-Ostriker-04,
Conroy-Wechsler-Kravtsov-06, Shankar-06, Conroy-Wechsler-Kravtsov-07, 
Baldry-Glazebrook-Driver-08, Moster-10, Guo-10, Neistein-11a, Neistein-11b, Yang-12}. 
The model assumes that each halo has a galaxy at its center and the stellar 
mass of a galaxy is an increasing function of the maximum mass ever attained 
by its halo. In practice, we obtain the relationship between dark halo mass $M_{200}$, 
as defined in \S~\ref{sec:simulations}, and the stellar mass $M_\ast$ 
of the central galaxy by matching the number density of dark matter halos 
with mass above $M_{200}$ with the number density of galaxies with 
stellar mass above $M_\ast$. For this we have used the stellar mass 
function of galaxies in the local Universe measured by \citet{Li-White-09}
from the SDSS/DR7 galaxy sample, corrected in \citet{Guo-10} using 
``model'' magnitudes. For the analysis in this subsection we assume there is 
no scatter in the $M_{200}-M_\ast$ relationship,
and we will take into account the effect of the scatter in later analyses. 

We note that, as pointed out recently by \citet{Yang-12}, one problem of the
current SHAMs is that they generally neglect the evolution of the central 
galaxy--host halo relation \citep[see also][]{Neistein-11b} and the different 
stripping/disruption of satellite galaxies with respect to the subhalos. 
Realistically,  satellite galaxies  
were accreted at higher redshifts and should have experienced different 
evolutionary processes from the centrals. Nevertheless, as shown in 
\citet{Yang-12}, current high redshift observations do imply that there
is little evolution of the central-host halo relation at {\em low} redshifts
 ($z\la 1$). Moreover, the central-host halo relation for halos with masses 
$\ga 10^{12}M_\odot$ is found to be insensitive to these caveats. Thus the 
approach of SHAMs is quite safe for our study where the galaxies are at 
$z\sim 0$ and are hosted by dark halos with mass well above $10^{12}M_\odot$ 
(see below).

The galaxy groups used here are  identified using
  the so called halo-based group  finder, where galaxies are grouped according
  to  their common  dark matter  halos following  the NFW  profiles,  i.e., by
  assuming the halos to be Spherical Overdensity (SO) structures. Such kind of
  SO  method  can  be  quite  different  from the  FoF  algorithm  where  each
  halo/group is  a self-bound virialized system.   On the other  hand, the FoF
  halos might have bridging effect ---  in some cases, two or more subhalos
  that are well  separated in space are associated with a  common FoF halo and
  only one of them (usually the most massive one) is classified as the central
  subhalo.  In order to {\em roughly} take into account the difference between 
  the FoF and SO halo finding algorithm, we have found out 
all the satellite subhalos that are located well beyond the virial radius 
of their central subhalo ($R_{200,c}$), by requiring that $d_{cs}>R_{200,c}+R_{200,s}$, 
where $d_{cs}$ is the distance from the central subhalo and $R_{200,s}$ 
is the virial radius of the satellite subhalo at the epoch when it was last
the central object of its host halo (see \S\ref{sec:simulations}).
We re-identify such satellite subhalos as central subhalos and include
them into the central galaxy subsamples when performing the analysis
below.

We have estimated both the cross-correlation functions and velocity 
dispersion profiles for the halo catalog. First, the central subhalos 
are divided into a set of subsamples according to stellar mass, with 
the same mass intervals as listed in Table 1 for the real subsamples. 
Next, each subsample is cross-correlated with the full catalog to compute 
a redshift-space cross-correlation function, $\xi^{(s)}(r_p,\pi)$, as
well as a projected cross-correlation function, $w_p(r_p)$. We have
followed the distant-observer approximation, defining the $x-y$ plane 
of the simulation box as the `sky' and the $z$-axis as the line of sight. 
The difference in the $z$-axis position between two galaxies and 
the difference in their peculiar velocity along the same axis combine to 
give a line-of-sight separation, $\pi$, thus including the redshift 
distortions in the computation of $\xi^{(s)}(r_p,\pi)$. We should point 
out that we have added a random component to the peculiar velocity of 
the central subhalos, assuming it to follow a Gaussian distribution 
function with a $1\sigma$ width of $0.25V_{200}$, where $V_{200}$ is 
the virial velocity of the halo. This takes into account the fact that 
the central galaxies of dark halos are not strictly at rest, but move 
relative to the halo center with a velocity that is 20-30\% the
virial velocity (\citealt{Diaferio-99, Berlind-03, Yoshikawa-Jing-Borner-03,
vandenBosch-05}; see also \citealt{Skibba-11} and references therein). 
Finally, the velocity dispersion profiles are determined by applying the 
$\xi^{(s)}(r_p,\pi)$ model in Eqn.~(\ref{eqn:xipvmodel}) to the 
$\xi^{(s)}(r_p,\pi)$ measured from the halo catalogs, in exactly the 
same way as adopted for the real samples. 

In Fig.~\ref{fig:sigma_6127} we plot in blue dash-dotted lines the 
velocity dispersion profiles measured from one of our simulations, the 
one with $\sigma_8=0.85$, and compare these with measurements from the 
SDSS data (symbols with error bars). At all masses, the model and the 
data show very similar profiles over the full range of $r_p$ probed.
In particular, the bump at around 1 Mpc is seen clearly in the model.
For systems with intermediate to high 
masses ($M_\ast\ga 5\times10^{10}M_\odot$), the bump in the model 
is comparable to that observed, in both amplitude and shape. 
At lower masses, the bump is stronger in the model.
As pointed out above, one possible reason behind the bump in the observed 
profiles is the contamination from neighboring halos in which the
peculiar motions of galaxies can contribute significantly to the velocity 
dispersion measurement of the system in question. To test this hypothesis,
we have done a simple experiment in which we exclude all the halos
from our central galaxy subsamples that are close to a neighboring larger 
halo, with the distance between the two halos smaller than the sum of 
their virial radii. The resulting velocity dispersion profiles are 
plotted in red solid lines in Fig.~\ref{fig:sigma_6127}. 
As can be seen, the bumps at around 1 Mpc become very weak or disappear,
as expected. 

The change in the velocity dispersion profile is strongest at the lowest 
masses where the bump is no longer
significant after the close halos are `blended'. We have played around
with different distance limits for blending the halos, and found that 
the bump becomes stronger for smaller blending radii. This can be 
understood from the fact that a smaller blending radius will include 
more satellite subhalos located in the inner region of FoF halos where 
the contamination in velocity is more serious.
At the highest masses ($M_\ast\ga 2\times10^{11}M_\odot$),
the velocity dispersion profile is essentially unchanged, reflecting 
the fact that the majority of the high-mass galaxies in the local
Universe are central galaxies of dark matter halos. At these masses,
the satellite fraction is found to be as low as $\sim 10\%$ by 
galaxy-galaxy lensing analyses of the SDSS data \citep{Mandelbaum-06},
as well as the SDSS-based halo occupation models \citep{Cooray-06,
Tinker-07b, vandenBosch-07}.

It is encouraging that the `deblending' effect doesn't change the
velocity dispersion profile within the virial radius for all the
stellar mass subsamples, except the lowest ones with 
$\log M_\ast/M_\odot\la 10.3$. This shows that, although the 
profile beyond the virial radius is somewhat uncertain due to the 
contamination from neighboring larger systems, the inner part can be 
well determined with the current data and methodology. Therefore, 
in what follows we limit our analyses to masses above 
$\log M_\ast/M_\odot=10.3$ and focus on velocity dispersions within 
the virial radius, and we use the original subhalos identified by
our HBT subhalo finder, without further mimicking the group finding
algorithm of \citet{Yang-07}. 

\subsection{Constraining density fluctuation parameter $\sigma_8$
and the stellar mass vs. halo mass relation}
\label{sec:sigma8}

\begin{figure*}
  \begin{center}
    \epsfig{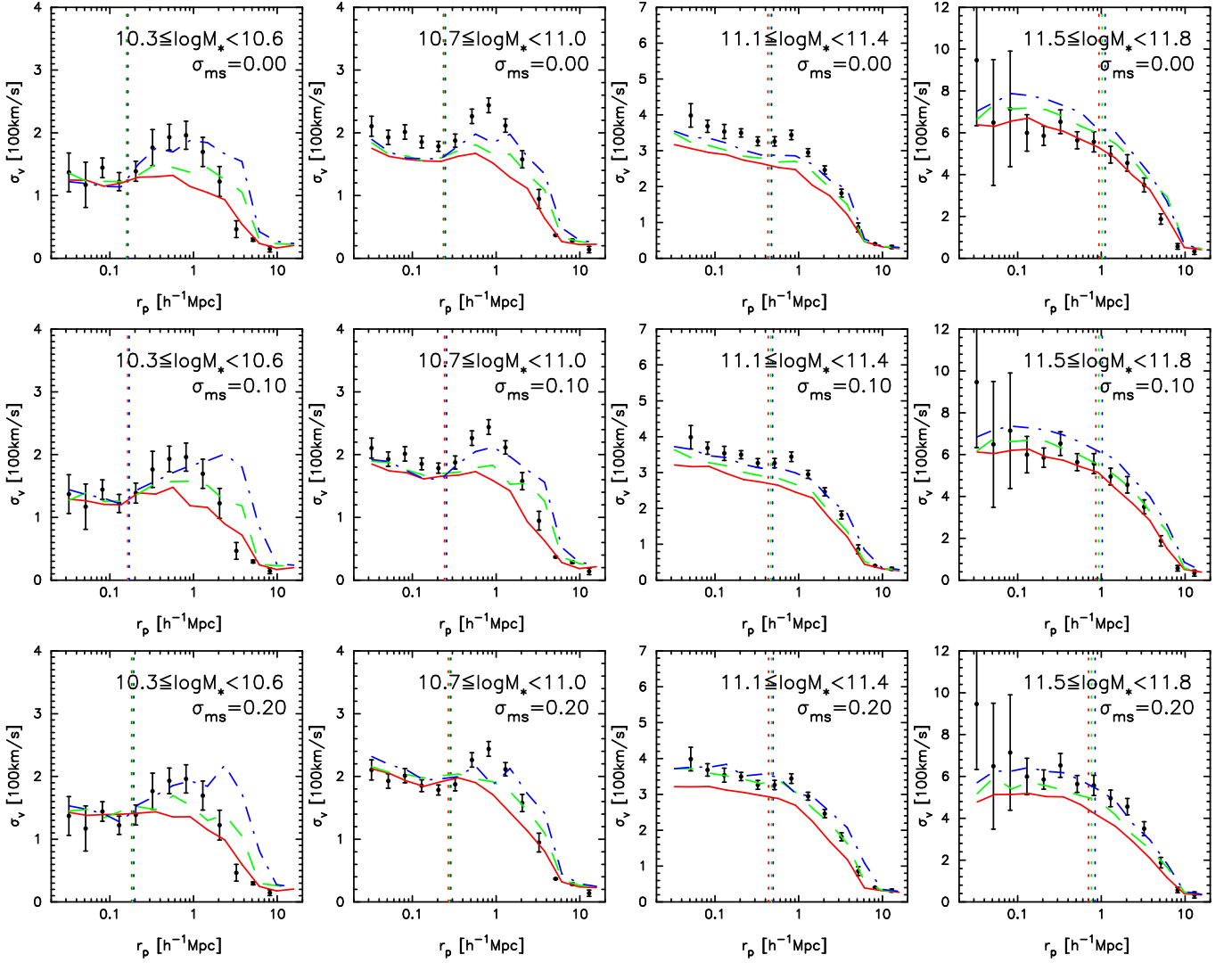}
  \end{center}
  \caption{Velocity dispersion profiles estimated from subhalo catalogs
for different stellar mass intervals (panels from left to right),
and for models with different dispersions in the stellar mass vs. halo
mass relation (panels from top to bottom) and with different density
fluctuation parameter: $\sigma_8$ = 0.75 (red solid line), 0.85 (green
dashed line) and 0.95 (blue dash-dotted line). The estimates
from the SDSS data are plotted in symbols with error bars for comparison.
The virial radii from the models are indicated by the vertical lines.}
  \label{fig:sigma_more}
\end{figure*}

The velocity dispersion of galaxies reflects the action of the local
gravitational field, and so should provide independent constraints
on the connection between galaxies and dark matter halos. In addition,
the relationship between galaxy velocity dispersion and mass is 
expected to depend on the density fluctuation parameter $\sigma_8$.
This is simply because $\sigma_8$ largely determines the abundance 
of the most massive halos, thus the velocity dispersion of the most
massive galaxies that are hosted by these halos. Here, we extend the
analysis in the previous subsection and show that the $\sigma_8$
parameter and the stellar mass vs. halo mass relation can be 
simultaneously constrained using our estimates of velocity dispersion 
obtained above.

To the end we use all the simulations with different $\sigma_8$ values. 
We also allow the central galaxy mass of halos of given dark matter
mass to have a certain amount of dispersion. Following the standard
practice, we assume the dispersion in $\log M_\ast$ to be Gaussian
and independent of halo mass, with rms values ranging from 
$\sigma_{ms}=0$ to 0.2 dex. The rms values exceeding about 0.2 dex 
are excluded because they are inconsistent with the steep high-mass 
falloff of the stellar mass function of \citet{Li-White-09}.

In Fig.~\ref{fig:sigma_more} we plot the velocity dispersion profiles
estimated from the subhalo catalogs for four of the stellar mass intervals 
(panels from left to right), and for models with different dispersions 
in the stellar mass vs. halo mass relation (from top to bottom) and with
different $\sigma_8$ values (the colorful lines). The observational
results are plotted in symbols with error bars for comparison. 
The virial radii are indicated in each panel with the vertical lines,
which are given by Eqn.(\ref{eqn:m200}) for the median halo mass of 
subhalos in a given stellar mass subsample. Overall, the velocity 
dispersions within the virial radii depend on both $\sigma_8$ and 
$\sigma_{ms}$, but the behavior differs with stellar mass. At low
masses, the velocity dispersion increases with increasing $\sigma_{ms}$,
but shows very little dependence on $\sigma_8$. At high masses, the
velocity dispersion depends on both parameters with a stronger trend
with $\sigma_8$. The velocity dispersion is higher for systems with
fixed mass and with fixed $\sigma_{ms}$ if $\sigma_8$ is larger, 
well consistent with the naive expectation as discussed above.
The weak dependence on $\sigma_{8}$ at the low-mass end is very
encouraging, in the sense that the degeneracy between $\sigma_8$ and 
$\sigma_{ms}$ as seen at high masses can be effectively broken,
allowing the two parameters to be constrained simultaneously.

\begin{figure*}
  \begin{center}
    \epsfig{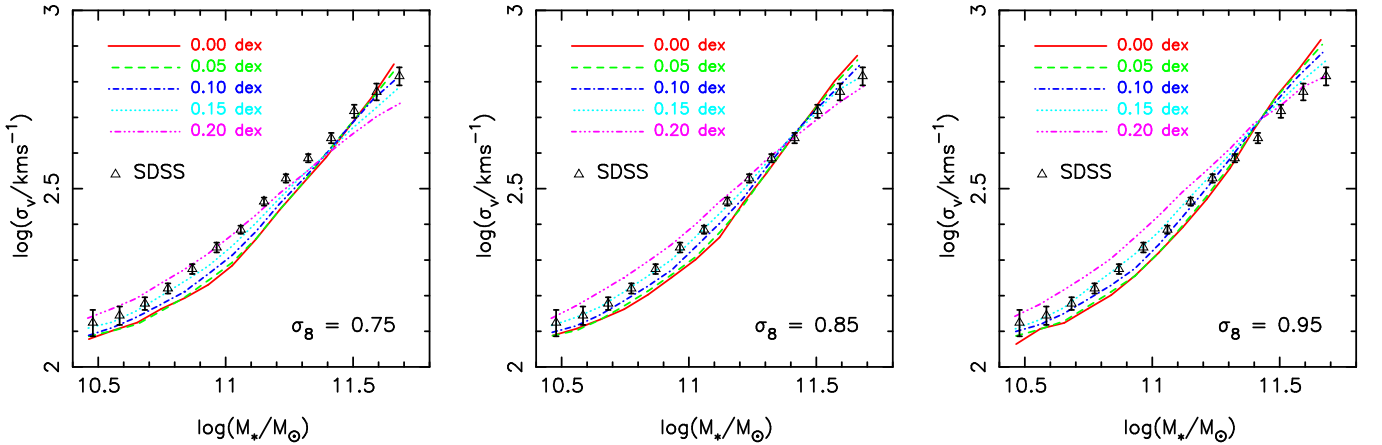}
  \end{center}
  \caption{Velocity dispersion averaged over $0.3R_{200}<r_p<R_{200}$
is plotted as a function of central galaxy mass. Results from the SDSS 
data are plotted in symbols with error bars and are repeated in every panel. 
Results from the models are plotted in different lines, which have different 
values of $\sigma_8$ (panels from left to right) and $\sigma_{ms}$ (different 
lines in each panel).}
  \label{fig:sm_sig}
\end{figure*}

For each stellar mass interval, we determined an average velocity
dispersion using the measurements falling in the range $0.3R_{200}<r_p<R_{200}$.
We exclude the inner-most part within $0.3R_{200}$ considering
the facts that the velocity dispersion profile slightly increases
at the smallest radii for massive systems (see Fig.~\ref{fig:profiles}) 
and this is not seen in the models at similar masses, and that 
we don't have velocity dispersion measurements on scales below $0.3R_{200}$
for low mass systems (again see Fig.~\ref{fig:profiles}). The former
fact is consistent with previous findings that the inner slope determined 
from gravitational lensing analyses is steeper than that expected from 
the NFW profile, implying that baryonic
processes have modified the inner profiles \citep[e.g.][]{Koopmans-06}. 
The average velocity dispersions obtained from the SDSS data are 
listed in Column 5 of Table 1. In Fig.~\ref{fig:sm_sig} we show the 
average velocity dispersion $\sigma_v$ 
as a function of stellar mass. Results from the SDSS data are plotted 
in symbols with error bars and are repeated in every panel. These 
are compared with results from the various models which have different 
values of $\sigma_8$ (panels from left to right) and $\sigma_{ms}$ 
(different lines in each panel). The degeneracy between $\sigma_8$ and 
$\sigma_{ms}$, as well as the capability of the $\sigma_v-M_\ast$ 
relation for breaking the degeneracy, are seen more clearly in this
figure. At low masses, the velocity dispersion shows strong dependence
on $\sigma_{ms}$. The dependence on $\sigma_{ms}$ is much weaker
at high masses, particularly at $\sim2\times10^{11}M_\odot$ where
the velocity dispersion is essentially determined by $\sigma_8$ 
alone. The data at these masses clearly suggest that an intermediate 
value of $\sigma_8=0.85$ works better than either higher or lower values.

\begin{figure}
  \begin{center}
    \epsfig{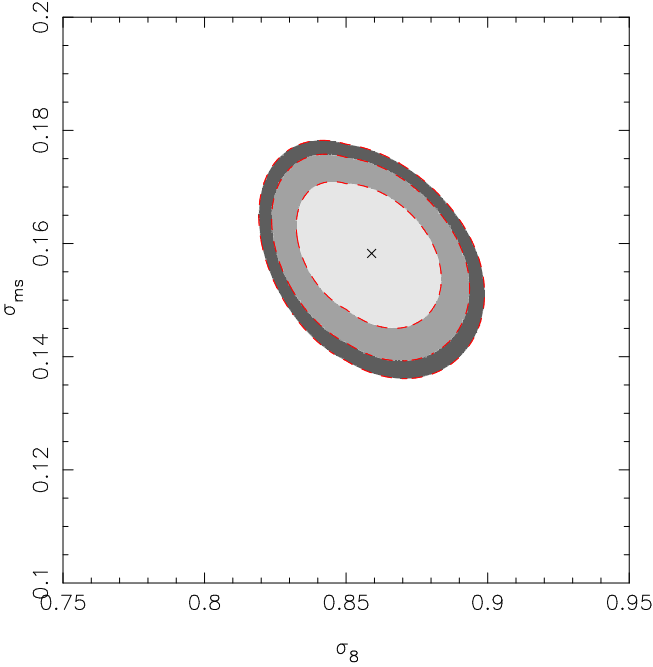}
  \end{center}
  \caption{$\chi^2$, defined by Eqn.(\ref{eqn:chi2}) and derived by
comparing the average velocity dispersion within the virial radius
of dark matter halos as predicted by the subhalo abundance matching 
model and as measured from groups of different central galaxy masses 
from the SDSS data, is plotted with respect to the minimum value 
$\chi^2_{min}$ (indicated by an `x'), in the grid of the two model 
parameters, $\sigma_8$ and $\sigma_{ms}$. The 68.3\%, 90\% and 95.4\%
confidence levels are plotted as red dashed lines, and the area
between every two neighboring levels is filled with shaded regions.}
  \label{fig:contours}
\end{figure}

In order to have a more quantitative constraint on both $\sigma_8$
and $\sigma_{ms}$, we have performed the analysis for more models
by varying the two parameters, with $\sigma_{ms}$ ranging from 0
to 0.2 dex with a step size of 0.01, and $\sigma_8$ ranging from
0.75 to 0.95 with a step size of 0.01. For given $\sigma_{ms}$,
we estimate velocity dispersions for models of different $\sigma_8$ 
by interpolating the $\sigma_v-M_\ast$ relation obtained from the
three simulations, assuming the velocity dispersion at fixed stellar 
mass to scale linearly with $\sigma_8$. This assumption is reasonably
true considering the following two facts. First, the change in velocity 
dispersion with $\sigma_8$ is fairly small, as can be seen from 
Fig.~\ref{fig:sm_sig}. Second, when obtaining the number density
of galaxy clusters through the \citet{Press-Schechter-74} formalism,
\citet[][see their Eqns. 7.75 and 7.76]{Mo-vandenBosch-White-10} found 
that $\sigma_8$ scales with cluster mass as 
$\sigma_8\propto M^{\beta/2}\Omega_m^{-\beta/3}$, 
where $\beta\approx 0.6
+0.8h\Omega_m$ for CDM-type power spectra with index $n=1$, leading
to a nearly linear relation between $\sigma_8$ and velocity dispersion
$\sigma_v$ for the assumed cosmology, $\sigma_v\propto\sigma_8^{0.9}$,
given that the mass is roughly proportional to $\sigma_v^3$.  

We compare the velocity dispersions
of each model to the SDSS measurements, and define the best-fitting 
model to be the one giving a minimum $\chi^2$ computed as follows:
\begin{equation}\label{eqn:chi2}
\chi^2\left(\sigma_8,\sigma_{ms}\right)=\mathbf{X^TC^{-1}X}.
\end{equation}
Here, $\mathbf{X}=\{X_j\}(j=1,...,m)$ is an $m\times1$ vector with
\begin{equation}
X_j=\sigma_{v,j}^{model}-\sigma_{v,j}^{SDSS},
\end{equation}
where $m$ is the number of stellar mass subsamples, $\sigma_{v,j}^{SDSS}$
is the velocity dispersion measured for the $j$th subsample from the 
SDSS and $\sigma_{v,j}^{model}$ is the result for the $j$th subsample
from the model. The $m\times m$ matrix $\mathbf{C}=\{C_{ij}\}(i,j=1,...,m)$
is the covariance matrix of the measurements from the 100 bootstrap
samples constructed from the SDSS data, given by
\begin{equation}
C_{ij} = \frac{1}{n}\left[ \sum_{k=1}^{n}\left(\sigma_{v,ki}^{boot}-\sigma_{v,i}^{SDSS}\right)
\left(\sigma_{v,kj}^{boot}-\sigma_{v,j}^{SDSS}\right) \right],
\end{equation}
where $n=100$ is the number of bootstrap samples, and $\sigma_{v,ki}^{boot}$ 
is the measurement for the $i$th stellar mass interval from the $k$th
bootstrap sample.

\begin{figure}
  \begin{center}
    \epsfig{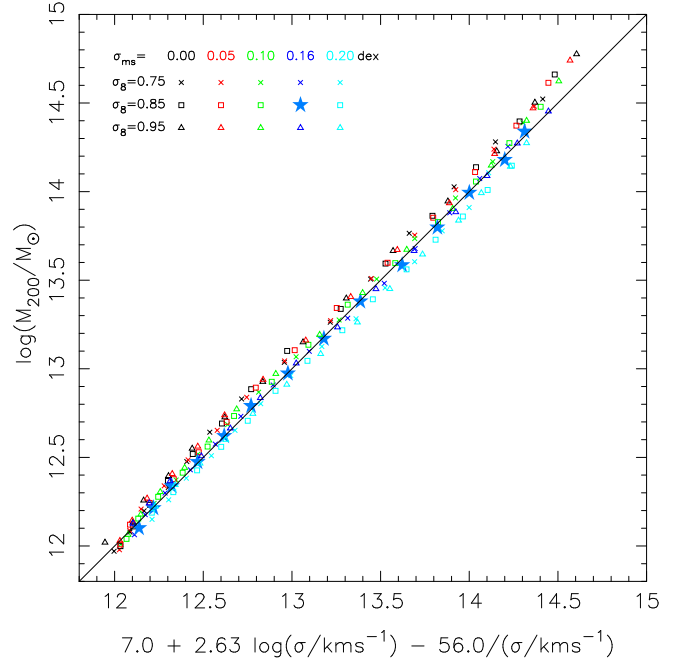}
  \end{center}
  \caption{Median halo mass $M_{200}$ for central subhalos with given
stellar mass is plotted as a function of the velocity dispersion 
measured using our methodology, for the models with different
values of $\sigma_8$ and $\sigma_{ms}$. The black solid line is the 
relation obtained by fitting to the result from the model with 
$\sigma_8=0.85$ and $\sigma_{ms}=0.15$ (plotted as big stars),
and is adopted as our calibration for estimating halo masses from
velocity dispersion measurements. The calibrated estimator is 
indicated at the bottom of the figure.}
  \label{fig:calibration}
\end{figure}

Fig.~\ref{fig:contours} plots the contours of $\Delta\chi^2=\chi^2-\chi^2_{min}$
in the grid of the two parameters. The 1, 2 and $3\sigma$ confidence
regions, computed for $m=14$ and 2 parameters, are indicated using
the dashed lines plus shaded regions. The minimum $\chi^2_{min}$ appears
at $\sigma_8=0.86$ and $\sigma_{ms}=0.16$ with $\chi^2/d.o.f=0.4$,
indicating that the fit is acceptable. There is a mild degeneracy between
the two parameters in the sense that the models with smaller $\sigma_8$
and larger $\sigma_{ms}$ and the models with larger $\sigma_8$ and
smaller $\sigma_{ms}$ can both provide a reasonable fit to the data.
While the minimum is the preferred solution at the $1\sigma$ level,
there is an elliptical degeneracy region extending to (0.82, 0.17) and 
(0.90, 0.14), that are allowed at $2-3\sigma$ level. Our constraint
on $\sigma_{ms}$ is in good agreement with previous studies which 
also found that the dispersion in stellar mass at fixed halo mass is 
$\sim0.13-0.20$ dex \citep[e.g.][]{Wang-06, Wang-07, Yang-Mo-vandenBosch-09,
Behroozi-Conroy-Wechsler-10, Moster-10, More-11, Yang-12}. 

Our constraint on $\sigma_8$ is slightly larger than the
result from the Wilkinson Microwave Anisotropy Probe 7 (WMAP7) and 
low redshift supernova and baryon acoustic oscillation data \citep{Komatsu-11},
which was $\sigma_8=0.816$. We note that this difference can be largely
explained by the slightly different matter density parameter, which was 
$\Omega_m=0.275$ in \citet{Komatsu-11} and is $\Omega_m=0.268$ in our 
simulations. This implies that our preferred value would be shifted
to $\sigma_8\approx 0.84$ for the WMAP7 matter density parameter, thus
within the $1\sigma$ level of the WMPAP7 result.

\subsection{Estimating galaxy halo masses}

\begin{figure*}
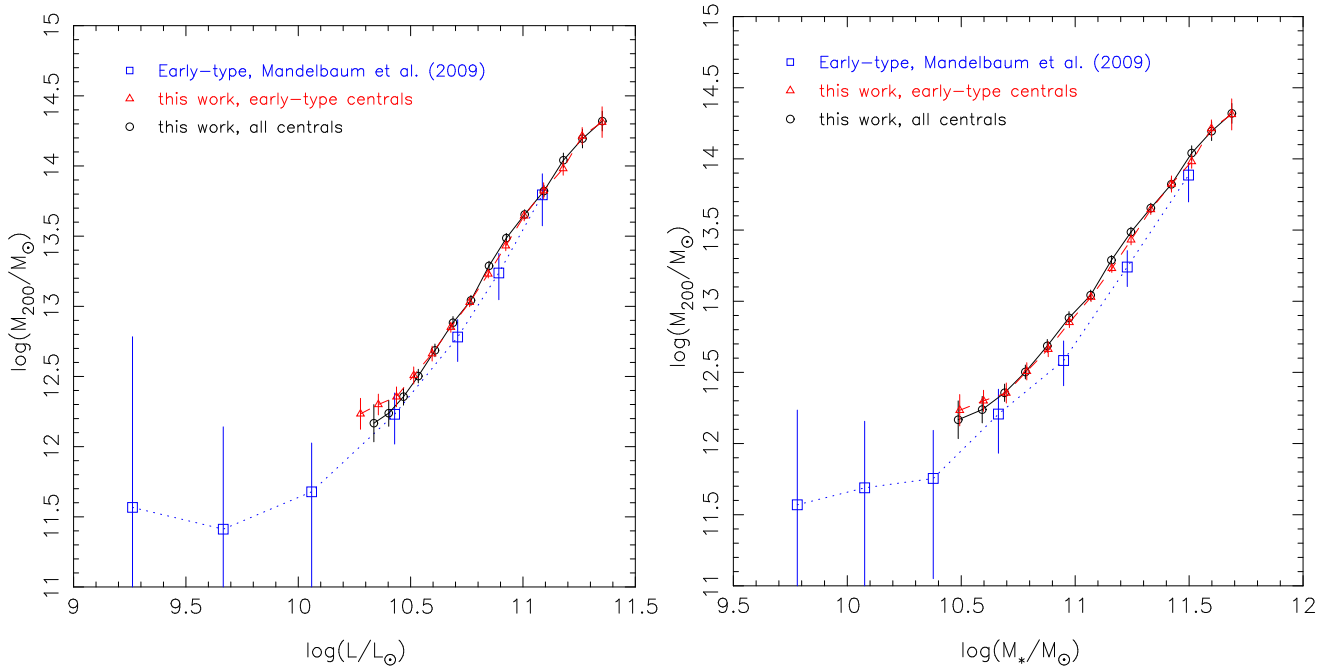

  \begin{center}
    \epsfig{figure=f10a.ps,width=0.48\hsize}
    \epsfig{figure=f10b.ps,width=0.48\hsize}
  \end{center}
  \caption{Dark matter halo mass as function of galaxy luminosity
  (left-hand panel) and stellar mass (right-hand panel). Black circles
  and red triangles are estimated from the velocity dispersion 
  measurements, for all the central galaxies in our group catalog
  and for the subset of early-type centrals, respectively.
  Blue squares are the result obtained by \citet{Mandelbaum-06} by
  stacking the gravitational lensing signals of the early-type
  central galaxies in an earlier SDSS data release.
  }
  \label{fig:sm_hm}
\end{figure*}

The velocity dispersion is caused by the local gravitational field
and so provides a direct measure of the dark halo mass of galaxies.
We use our simulations to calibrate the relationship between the 
velocity dispersion measured using our methodology and the halo mass 
for central subhalos of different stellar masses. As plotted in 
Fig.~\ref{fig:calibration}, the best-fitting relation is 
\begin{eqnarray}\label{eqn:estimator}
\log (M_{200}/M_\odot) & = & 7.0+2.63\log (\sigma_v/\mbox{kms}^{-1}) \nonumber\\
                           &   & -56.0/(\sigma_v/\mbox{kms}^{-1}),
\end{eqnarray}
obtained using the subhalo abundance matching model 
applied to the simulation with $\sigma_8=0.85$ and with the dispersion
parameter $\sigma_{ms}=0.16$.  As shown above, this model provides
the best match to the SDSS data in the stellar mass versus velocity
dispersion relation. 

We apply the halo mass estimator in Eqn.(\ref{eqn:estimator}) to our
measurements of velocity dispersions. The resulting halo masses are
listed in the last column of Table 1 and are plotted in Fig.~\ref{fig:sm_hm} 
as a function of both luminosity
(left-hand panel) and stellar mass (right-hand panel). For comparison,
we also plot the results obtained by \citet{Mandelbaum-06} from stacking 
the gravitational lensing signals from the SDSS data. We note that
about 90\% of our central galaxies are {\em early-type} galaxies, 
which are defined in practice to have the parameter {\tt frac\_deV} $\ge0.5$
from the SDSS {\sc photo} pipeline. In order to make a
fair comparison, we have measured the velocity dispersions for groups
with early-type centrals and plot the results as red triangles in 
Fig.~\ref{fig:sm_hm}. The result doesn't change much when compared to
that of the full sample. The halo masses of early-type centrals presented
in \citet{Mandelbaum-06} are plotted as open squares. As can be
seen, the halo masses estimated from galaxy velocity dispersions agree 
pretty well with those from gravitational lensing signals, over the
full luminosity range ($2\times10^{10}L_\odot\la L\la1.5\times10^{11}M_\odot$) 
where the halo mass estimates are available from both studies. Note 
that we have shifted the halo masses of \citet{Mandelbaum-06} by about 
-0.1 dex in order to take into account the different halo mass definitions
adopted in the two studies.

The agreement becomes slightly worse when the halo mass is plotted as 
a function of stellar mass, as shown in the right-hand panel of 
Fig.~\ref{fig:sm_hm}. In this case our halo mass is systematically 
higher than that from \citet{Mandelbaum-06} by about 50\% at 
intermediate to low masses. The stellar masses used in \citet{Mandelbaum-06} 
were estimated by \citet{Kauffmann-03} based on the SDSS spectroscopy, 
while those in this study are based on the photometric properties. 
Therefore, for this comparison in Fig.~\ref{fig:sm_hm}, we have shifted the 
former by about 0.1 dex in stellar mass; the correction factor was obtained 
in Appendix A of \citet{Li-White-09} using a large sample of galaxies for which 
both stellar mass estimates are available 
\citep[see also fig.17 of][]{Blanton-Roweis-07}. As can be seen 
in \citet{Li-White-09}, the difference between the two stellar masses 
depends on stellar mass in a complicated way with varying scatter.
Thus, a constant correction cannot fully get rid of the difference,
and the larger discrepancy between our results and those of \citet{Mandelbaum-06}
as expressed in terms of stellar mass must be partially (if not totally) 
due to this effect.

\section{Summary and discussion}

Using a large catalog of galaxy groups selected from the Sloan Digital 
Sky Survey (SDSS) data release 7 (DR7) by \citet{Yang-07}, we have 
derived the velocity dispersion profiles for groups with different 
masses of central galaxies, by modelling
the redshift distortions in group-galaxy cross-correlation function. 
Our main results can be summarized as follows.
\begin{itemize}
\item An \citet{Navarro-Frenk-White-97} profile on small
scales plus a biased version of the linear mass autocorrelation function
can well describe the observed $w_p(r_p)$ over all the scales probed
($15\mbox{kpc}<r_p<30\mbox{Mpc}$), thus providing an accurate determination 
of the real-space cross-correlation function $\xi_{cg}(r)$. This has been
well expected from theoretical studies of halo-mass and galaxy-mass
cross-correlation functions \citep[e.g.][]{Hayashi-White-08}.
\item The velocity dispersion within virial radius ($R_{200}$)
shows a roughly flat profile, with a slight increase at the smallest radii
($\la 0.3 R_{200}$). This is consistent with the theoretical expectation
that baryonic condensation has modified the inner profile.
\item The average velocity dispersion within the virial radius, $\sigma_v$,
is a strongly increasing function of central galaxy mass, ranging from
$\sim$130 kms$^{-1}$ for the lowest-mass groups in our sample which have
central galaxy masses of $M_\ast\sim3\times10^{10}M_\odot$, up to
$\sim$650 kms$^{-1}$ for the highest-mass systems with
$M_\ast\sim5\times10^{11}M_\odot$. The mass dependence is more remarkable
 at higher masses.
\end{itemize}

We have extended our analysis of the SDSS data to a set of $N$-body simulations 
with the concordance $\Lambda$ cold dark matter ($\Lambda$CDM) cosmology but
different values of the density fluctuation parameter $\sigma_8$.
We have assigned a stellar mass to each of our subhalos under the plausible 
hypothesis that the stellar mass of central galaxies is an increasing 
function of the dark matter mass of their host halos, usually called subhalo
abundance matching model in the literature, and measured the velocity dispersion 
profile for the subhalos in different stellar mass intervals. By comparing 
the results to measurements from the SDSS data, we have shown that 
 the $\sigma_v-M_\ast$ relation obtained from the data provides
stringent constraints on both the $\sigma_8$ parameter and $\sigma_{ms}$,
the dispersion in $\log M_\ast$ of central galaxies at fixed halo mass.
Our best-fitting model suggests that $\sigma_8=0.86\pm0.03$ and
$\sigma_{ms}=0.16\pm0.03$. The latter is in very good agreement with 
previous studies where $\sigma_{ms}$ was inferred from semi-analytic catalogs 
of galaxy formation models \citep[e.g.][]{Wang-06, Wang-07}, from group catalogs 
\citep{Yang-Mo-vandenBosch-09}, from satellite kinematics analysis \citep{More-11}, 
from clustering analysis \citep{Moster-10}, and from subhalo abundance matching 
\citep{Behroozi-Conroy-Wechsler-10, Guo-10, Yang-12}. 

Our velocity dispersion measurements also provide a direct measure 
of the dark matter mass of the host halos for central galaxies of different 
luminosities and stellar masses, which are in good agreement with the results 
obtained by \citet{Mandelbaum-06} from stacking the gravitational lensing 
signals of the SDSS galaxies. The results also broadly agree with recent 
estimates by many other authors \citep[e.g.][]{vandenBosch-04,Yang-07, 
More-09, Guo-10, More-11}, though with varying degrees of subtle discrepancies 
due to different methodologies and date sets.

It is encouraging to see that the velocity dispersion as a function of
central galaxy mass as measured in our study is able to constrain the
density fluctuation parameter $\sigma_8$, although the preferred value
is slightly higher than that from the most recent result from WMAP7
\citep{Komatsu-11}. We suspect that this difference can be explained if
the effect of the different matter density parameters adopted in the two 
studies is considered. It has been recently emphasized that the redshift-space
distortion on very large scales (near or beyond the baryon acoustic 
oscillation scale), expected to be well detected in next-generation
redshift survey like the BigBOSS experiment \citep{Schlegel-11}, will
provide powerful constraints on the growth factor of structure and 
the physical properties of dark energy \citep[e.g.][]{Okumura-Jing-11,
Samushia-Percival-Raccanelli-12}.  Our study 
suggests that the redshift distortions on galaxy and group scales 
provide complementary information useful for constraining cosmological
parameters. When combined with more observational measurements like
the autocorrelation function and pairwise velocity dispersion of galaxies 
as functions of their physical properties \citep[e.g.][]{Zehavi-05, Li-06a,Li-06c},
our results should be able to provide more accurate and reliable
constraints on $\sigma_8$. We will come back to this point in future
studies.

There is still room to improve our study. First, we have assumed isotropic
distributions for the peculiar velocity and velocity dispersion of
satellite galaxies and this is roughly, but not exactly correct. Models
incorporated with radial anisotropy might provide better determinations
of the velocity dispersions. Using red, early-type galaxies instead of
the whole population as the tracer may also help, as they are more likely 
consistent with isotropic orbits than are blue and late-type galaxies
\citep[e.g.][]{Carlberg-97a, Biviano-Katgert-04}. 
Second, we have used groups with three or more member galaxies and assumed 
that our determinations of velocity dispersion are insensitive to the 
richness limit of the group sample. Larger and deeper surveys in future 
would allow us to test this assumption and improve our method.
Third, previous studies found that Brightest Cluster Galaxies (BCG) may
be not at the true center of their host halo \citep[e.g.][]{Skibba-11},
and this effect could also introduce some sort of systematics.
Fortunately, this is unlikely a significant issue for us, as such offsets 
seem to be small in general (a recent study by \citealt{vonderLinden-12}
found a median value of 20 kpc in the ffsets between X-ray centroids and 
BCGs in clusters). Fourth, we have assumed the dispersion in
stellar mass is independent of halo mass and this might be tested and
improved via semi-analytic galaxy formation models or halo occupation
distribution models. Finally, our work is limited to relatively large
systems with $M_\ast\ga 3\times10^{10}M_\odot$ and could be extended
to lower masses using deeper surveys. Previous studies of the pairwise
velocity dispersion (PVD) of galaxies revealed that the relative velocities
between galaxies exhibit a well-defined minimum at intermediate luminosities/masses,
and both fainter and brighter galaxies have very high velocities, 
indicative of massive halos of cluster size \citep{Jing-Borner-04a, Li-06a}. 
It would be interesting to directly measure the halo mass for such faint 
galaxies by applying our methodology to groups of lower masses,  e.g. 
down to $\sim6\times10^{9}M_\odot$ using the Galaxy And Mass Assembly 
survey \citep[GAMA;][]{Driver-09}.

\acknowledgments

We are grateful to the referee for helpful comments.
CL is grateful to Qi Guo, Donghai Zhao, Simon White, Raul Angulo and
Ying Zu for helpful discussion, and acknowledges the support of the 
100-Talent Program of Chinese Academy
of Sciences (CAS), Shanghai  Pujiang Programme (no. 11PJ1411600)
and the exchange program between Max Planck Society and CAS.
This work is  sponsored  by NSFC  (11173045, 10878001, 11033006,
11121062, 10925314, 11128306) and the CAS/SAFEA International Partnership 
Program for  Creative Research Teams  (KJCX2-YW-T23).

Funding for  the SDSS and SDSS-II  has been provided by  the Alfred P.
Sloan Foundation, the Participating Institutions, the National Science
Foundation, the  U.S.  Department of Energy,  the National Aeronautics
and Space Administration, the  Japanese Monbukagakusho, the Max Planck
Society,  and the Higher  Education Funding  Council for  England. The
SDSS Web  Site is  http://www.sdss.org/.  The SDSS  is managed  by the
Astrophysical    Research    Consortium    for    the    Participating
Institutions. The  Participating Institutions are  the American Museum
of  Natural History,  Astrophysical Institute  Potsdam,  University of
Basel,  University  of  Cambridge,  Case Western  Reserve  University,
University of Chicago, Drexel  University, Fermilab, the Institute for
Advanced   Study,  the  Japan   Participation  Group,   Johns  Hopkins
University, the  Joint Institute  for Nuclear Astrophysics,  the Kavli
Institute  for   Particle  Astrophysics  and   Cosmology,  the  Korean
Scientist Group, the Chinese  Academy of Sciences (LAMOST), Los Alamos
National  Laboratory, the  Max-Planck-Institute for  Astronomy (MPIA),
the  Max-Planck-Institute  for Astrophysics  (MPA),  New Mexico  State
University,   Ohio  State   University,   University  of   Pittsburgh,
University  of  Portsmouth, Princeton  University,  the United  States
Naval Observatory, and the University of Washington.

\bibliography{ref}


\label{lastpage}
\end{document}